\documentclass[twocolumn,showpacs,preprintnumbers,amsmath,amssymb,pre]{revtex4}
\usepackage{graphicx}
\usepackage{dcolumn}
\usepackage{bm}

\begin{document}

\title{Phase diagram of self-assembled rigid rods
on two-dimensional lattices: Theory and Monte Carlo simulations}

\author{L. G. L\'opez}
\author{D. H. Linares}
\author{A. J. Ramirez-Pastor}
\email{antorami@unsl.edu.ar} \affiliation{Departamento de
F\'{\i}sica, Instituto de F\'{\i}sica Aplicada, Universidad
Nacional de San Luis, CONICET, 5700 San Luis, Argentina}

\author{S. A. Cannas}
\affiliation{Facultad de Matem\'atica, Astronom\'{\i}a y
F\'{\i}sica, Universidad Nacional de C\'ordoba and \\ Instituto de
F\'{\i}sica Enrique Gaviola  (IFEG-CONICET), Ciudad Universitaria,
5000 C\'ordoba, Argentina}

\date{\today}

\begin{abstract}

Monte Carlo simulations and finite-size scaling analysis have been
carried out to study the critical behavior in a two-dimensional
system of particles with two bonding sites that, by decreasing
temperature or increasing density, polymerize reversibly into
chains with discrete orientational degrees of freedom and, at the
same time, undergo a continuous isotropic-nematic (IN) transition.
A complete phase diagram was obtained as a function of temperature
and density. The numerical results were compared with Mean Field
(MF) and Real Space Renormalization Group (RSRG) analytical
predictions about the IN transformation. While the RSRG approach
supports the continuous nature of the transition, the MF solution
predicts a first-order transition line and a tricritical point, at
variance with the simulation results.

\end{abstract}

\pacs{05.50.+q, 
64.70.mf, 
61.20.Ja, 
64.75.Yz, 
75.40.Mg} 

\maketitle

\section{Introduction}

Molecular self-assembly is one of the basic mechanisms of life and
matter, and thus, modeling and measurements of naturally occurring
self-assembling systems has long been pursued in the biological
and physical sciences \cite{Pelesko,Krasnogor}. Despite the large
number of papers that are currently reported, many of the ideas
that are crucial to the development of this area (molecular shape,
interplay between enthalpy and entropy, nature of the forces that
connect the particles in self-assembled molecular aggregates) are
simply not yet under the control of investigators.

Self-assembly also poses a number of substantial technological
challenges
\cite{Nalwa,Whitesides,Jiang,Anderson,Gooding,Glotzer,Glotzer1,Love,Zaccarelli,Alberts,Lyons}.
In fact, the biological systems use self-assembly to assemble
macromolecules and structures. Imitating these strategies and
creating novel molecules with the ability to self-assemble into
supramolecular assemblies is an important technique in
nanotechnology. There is then a need for understanding the basic
principles governing this type of organization.

It is obvious that a complete analysis of the self-assembly
phenomenon is a quite difficult subject because of the complexity
of the involved microscopic mechanisms. For this reason, the
understanding of simple models with increasing complexity might be
a help and a guide to establish a general framework for the study
of this kind of systems, and to stimulate the development of more
sophisticated models which can be able to reproduce concrete
experimental situations.

Computer simulations have shown that spherical particles
interacting isotropically through repulsive interparticle
interactions can spontaneously assemble into anisotropic
structures \cite{Likos,Mladek,Malescio}. The presence of an
isotropic short ranged interparticle attraction coupled to a
longer ranged repulsion can also yield anisotropic structures.
However, most real components, from proteins to ions \cite{De
Yoreo} to the wide variety of recently synthesized nanoparticles
\cite{Glotzer,Glotzer1}, interact via anisotropic or ``patchy"
attractions. Simulation work \cite{Duff,Rein,Gee,Doye,Auer}
reveals assembly pathways of such components to be in general
richer than those of their isotropic counterparts. Experimental
realization of such systems is growing. An example of real patchy
particles is presented in Ref. [\onlinecite{Cho}]. Such particles
offer the possibility to be used as building blocks of
specifically designed self-assembled structures
\cite{Whitesides,Glotzer,Glotzer1,Zhang1,Zhang2}. Moreover, the
implications of patchy colloids for proteins \cite{Doye}, which
are patchy by nature, could be significant.

In this line, we consider in this paper the general problem of
particles with strongly anisotropic, highly directional
interactions in which effectively attractive patches induce the
reversible self-assembly of particles into chains, i.e.,
equilibrium polymerization
\cite{Sciortino,Tavares,Tavares1,PRE4,Tavares2,Tavares3,JCP13}.
Recently, several research groups reported on the assembly of
colloidal particles in linear chains. Selectively functionalizing
the ends of hydrophilic nanorods with hydrophobic polymers, Nie et
al. reported the observation of rings, bundles, chains, and
bundled chains \cite{Nie}. In another experimental study carried
out by Chang et al. \cite{Chang}, gold nanorods were assembled
into linear chains using a biomolecular recognition system. In a
direct relation with the present work, Clair et al. \cite{Clair}
investigated the self-assembly of terephthalic acid (TPA)
molecules on the Au(111) surface. Using scanning tunneling
microscopy, the authors showed that the TPA molecules arrange in
one-dimensional chains with a discrete number of orientations
relative to the substrate.

It is well known that solutions of self-assembled chains exhibit a
transition from a disordered isotropic phase to an ordered nematic
phase as the concentration of particles increases. Experimental
examples of equilibrium polymer systems that exhibit a
isotropic-nematic (IN) phase transition include wormlike micelles
\cite{Hoffmann} and self-assembled protein fibers like $f$-actin
\cite{Oda,Viamontes}. In this context, a recent paper was devoted
to the study of a system of self-assembled rigid rods adsorbed on
a two-dimensional lattice \cite{Tavares}. In Ref.
[\onlinecite{Tavares}], Tavares et al. studied a system composed
of monomers with two attractive (sticky) poles that polymerize
reversibly into polydisperse chains and, at the same time, undergo
a isotropic-nematic (IN) continuous phase transition \cite{foot0}.
So, the interplay between the self-assembly process and the
nematic ordering is a distinctive characteristic of these systems.
Using an approach in the spirit of the Zwanzig model
\cite{Zwanzig}, the authors found that nematic ordering enhances
bonding. In addition, the average rod length was described
quantitatively in both phases, while the location of the ordering
transition, which was found to be continuous, was predicted
semiquantitatively by the theory. With respect to the
characteristics of the phase transition, it has recently been
shown that, at intermediate density, the IN transition is in the
$q=1$ Potts universality class \cite{PRE4}.

\begin{figure}[t]
\includegraphics[width=6cm,clip=true]{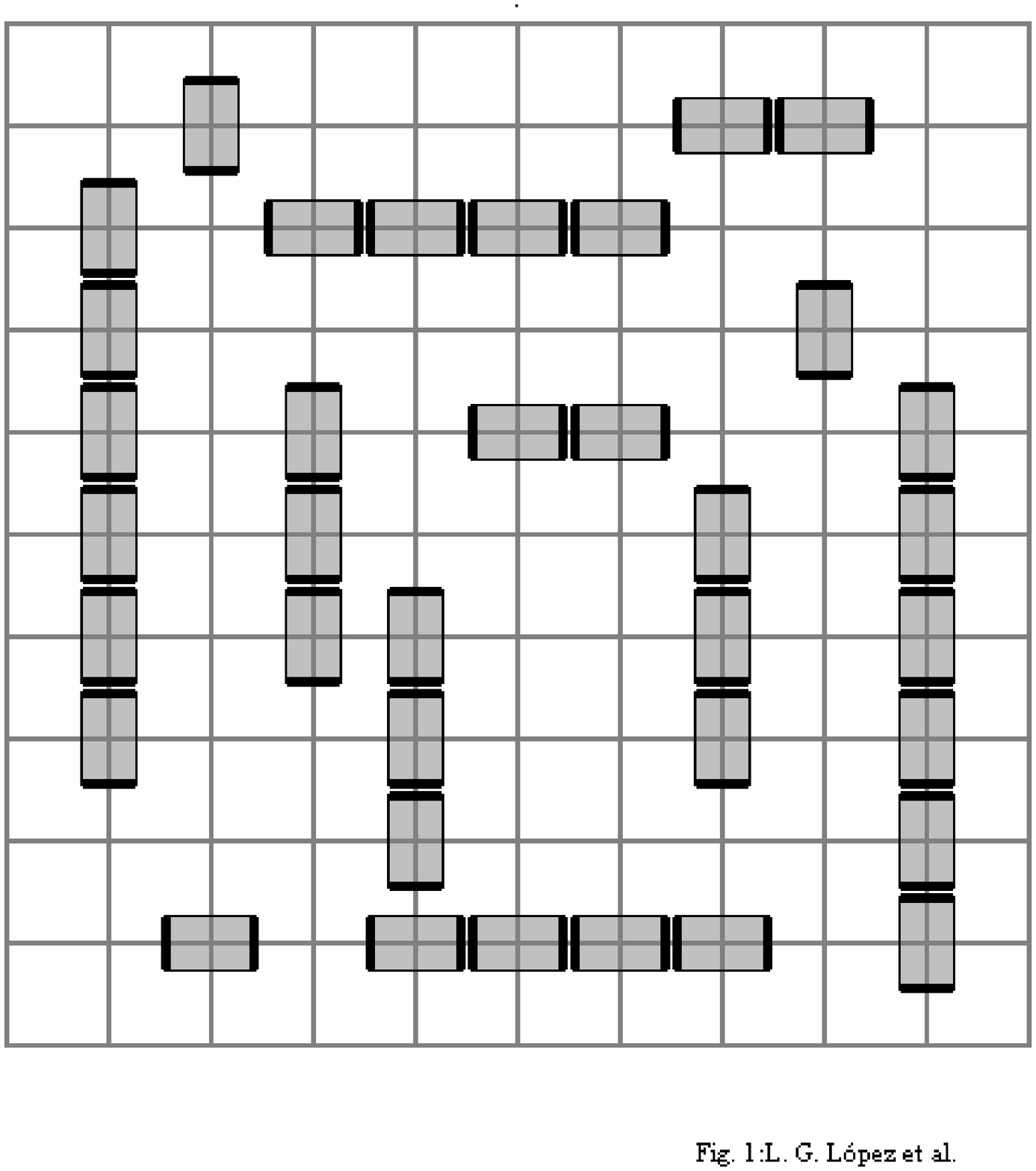}
\caption{Schematic representation of a system of self-assembled
rigid rods on a square lattice. } \label{figure1}
\end{figure}

The temperature-coverage phase diagram obtained in Ref.
[\onlinecite{Tavares}] is qualitative only, and the theory
overestimates the critical temperature in all range of coverage.
In addition, the possibility of a reentrant nematic transition at
high densities \cite{Ghosh} was not investigated  by Tavares et
al.. Accordingly, the main objective of the present work is to
provide an accurate determination of the phase diagram of the
system. For this purpose, extensive Monte Carlo (MC) simulations,
supplemented by finite-size scaling analysis and two analytical
approximations, have been carried out to obtain the critical
temperature characterizing the IN phase as a function of the
coverage. The paper is organized as follows. In Sec. II we
describe the lattice-gas model. The simulation scheme and
computational results are given in Sec. III. In Sec. IV we present
the analytical approximations [Mean-Field approximation (MF), and
Real Space Renormalization Group approach (RSRG)] and compare the
MC results with the theoretical calculations. Finally, the general
conclusions are drawn in Sec. V.

\section{Lattice-Gas Model}
\label{Lattice-Gas Model}

As in Refs. [\onlinecite{Tavares,PRE4}], we consider a system of
self-assembled rods with a discrete number of orientations in two
dimensions. We assume that the substrate is represented by a
square lattice of $M = L \times L$ adsorption sites, with periodic
boundary conditions. $N$ particles are adsorbed on the substrate
with two possible orientations along the principal axis of the
square lattice. These particles interact with nearest-neighbors
(NN) through anisotropic attractive interactions (see Fig. 1).
Then, the adsorbed phase is characterized by the Hamiltonian

\begin{equation}
H = -w \sum_{\langle i,j \rangle} |\vec{r}_{ij} \cdot
\vec{\sigma}_i||\vec{r}_{ji} \cdot \vec{\sigma}_j|, \label{h}
\end{equation}
where $\langle i,j \rangle$ indicates a sum over NN sites; $w$
represents the NN lateral interaction between two neighboring $i$
and $j$, which are aligned with each other and with the
intermolecular vector $\vec{r}_{ij}$; and $\vec{\sigma}_i$ is the
occupation vector with $\vec{\sigma}_i=0$ if the site $i$ is
empty, $\vec{\sigma}_i= \hat{x}$ if the site $i$ is occupied by a
particle with orientation along the $x$-axis, and $\vec{\sigma}_i=
\hat{y}$ if the site $i$ is occupied by a particle with
orientation along the $y$-axis.

A cluster or uninterrupted sequence of bonded particles is a
self-assembled rod. At fixed temperature, the average rod length
increases as the density increases and the polydisperse rods will
undergo an nematic ordering transition \cite{Tavares}.

Since each site state is characterized by a three-state variable,
we can rewrite Hamiltonian (\ref{h}) in terms of new variables
$S_i=0,\pm 1$, where where $S_i\pm 1$ represent the vertical
($\vec{\sigma}_i= \hat{y}$) and the horizontal ($\vec{\sigma}_i=
\hat{x}$) orientations, while $S_i=0$ represents the empty state.
Then, Hamiltonian (\ref{h}) reads

\begin{widetext}
\begin{eqnarray}
     H &=& -\frac{w}{4} \sum_{<i,j>} S_i S_j \left[ (S_i+1)(S_j+1)(\hat{y} . \vec{r}_{ij})+  (S_i-1)(S_j-1)(\hat{x}.\vec{r}_{ij}) \right] \nonumber \\
     &=& -\frac{w}{4} \sum_{<i,j>} \left[(S_i^2+S_i)(S_j^2+S_j)(\hat{y} . \vec{r}_{ij})+  (S_i^2-S_i)(S_j^2-S_j)(\hat{x}.\vec{r}_{ij}) \right]  \label{H3}
\end{eqnarray}
\end{widetext}

\noindent This Hamiltonian has the same energy spectrum as
(\ref{h}). Notice that the transformation $S_i \to -S_i$ is not a
symmetry of the Hamiltonian (\ref{H3}), since it is equivalent to
a $90^o$ rotation of the lattice, but it is a symmetry of the
system, since it left the partition function unchanged. The total
number of adsorbed particles can be written

\begin{equation}\label{Ns}
    N = \sum_i  S_i^2
\end{equation}

When $N=M$, we have $S_i^2=1$ and the Hamiltonian (\ref{H3})
results

\begin{equation}
    H(\{S_i\}) = -\frac{w}{4} \sum_{<i,j>} \left[1+S_i S_j + (S_i+S_j)(\hat{y}.\vec{r}_{ij}-\hat{x}.\vec{r}_{ij}) \right]
\end{equation}

The last sum in Eq.(4) vanishes and therefore $H_I = -\frac{w}{4}
\sum_{<i,j>} S_i S_j +$ constant.  Hence, in that limit the
present model reduces to the Ising one with coupling constant
$w^{Ising} = w/4$.

\section{Monte Carlo Simulations}

\begin{figure}[t]
\includegraphics[width=6cm,clip=true]{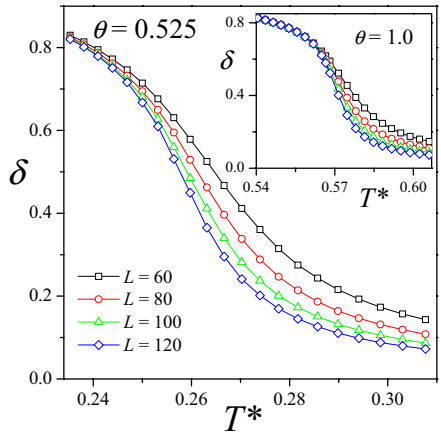}
\caption{Size dependence of the order parameter as a function of
temperature for $\theta= 0.525$ and $\theta = 1$ (inset). }
\label{figure2}
\end{figure}

\subsection{Monte Carlo Method}

We have used a standard importance sampling MC method in the
canonical ensemble \cite{BINDER} and finite-size scaling
techniques \cite{PRIVMAN}. The procedure is as follows. Starting
with a random initial configuration (sites occupied with
concentration $\theta=N/M$ and particle axis orientation chosen
with probability $1/2$), successive configurations are generated
by attempting to move single particles (monomers). One of the two
(translation or rotation) moves is chosen at random. In a
translation move, an occupied site and an empty site are randomly
selected and their coordinates are established. Then, an attempt
is made to interchange its occupancy state with probability given
by the Metropolis rule \cite{Metropolis}: $P = \min
\left\{1,\exp\left( - \beta \Delta H \right) \right\}$, where
$\Delta H$ is the difference between the Hamiltonians of the final
and initial states and $\beta=1/k_BT$ (being $k_B$ the Boltzmann
constant). For a rotation move, the rotational state of the chosen
particle (horizontal or vertical) is changed with a probability
determined by Metropolis criteria.

A Monte Carlo step (MCS) is achieved when $\theta \times M$ sites
have been tested to change its occupancy state. Typically, the
equilibrium state can be well reproduced after discarding the
first $5 \times 10^6$ MCS. Then, the next $6 \times 10^8$ MCS are
used to compute averages. All calculations were carried out using
the parallel cluster BACO of Universidad Nacional de San Luis,
Argentina. This facility consists of 60 PCs each with a 3.0 GHz
Pentium-4 processor and 90 PCs each with a 2.4 GHz Intel Core 2
Quad processor.

In order to follow the formation of the nematic phase from the
isotropic phase, we use the order parameter defined in Ref.
[\onlinecite{PRE4}],
\begin{equation}
\delta =  \frac{\left | N_v -N_h \right |}{N_v +N_h},
 \label{fi}
\end{equation}
where $N_h(N_v)$ is the number of monomers aligned along the
horizontal (vertical) direction ($N=N_h + N_v$).

In our Monte Carlo simulations, we set the density $\theta$,
varied the temperature $T$ and monitored the order parameter
$\delta$, which can be calculated as simple averages. The reduced
fourth-order cumulant, $U_L$, introduced by Binder \cite{BINDER},
was calculated as:
\begin{equation}
U_L(T) = 1 -\frac{\langle \delta^4\rangle} {3\langle
\delta^2\rangle^2}, \label{cum}
\end{equation}
\noindent where the thermal average  $\langle ... \rangle$, in all
the quantities, means the time average throughout the MC
simulation.

\subsection{Computational results}

The critical behavior of the present model has been investigated
by means of the computational scheme described in the previous
section and finite-size scaling analysis \cite{BINDER,PRIVMAN}.

We start with the calculation of the order parameter plotted
versus the reduced temperature $T^*=k_B T/ w$ for several lattice
sizes ($L = 60, 80, 100$ and $120$) and two values of coverage
[$\theta=0.525$ \cite{foot1}, Fig. 2 and $\theta=1$, inset of Fig.
2]. As it can be observed, $\delta$ appears as a proper order
parameter to elucidate the phase transition. When the system is
disordered ($T^*>T^*_c$, being $T^*_c$ the critical temperature),
all orientations are equivalents and $\delta$ is zero. In the
critical regime ($T^*<T^*_c$), the particles align along one
direction and $\delta$ is different from zero.

\begin{figure}[t]
\includegraphics[width=5cm,clip=true]{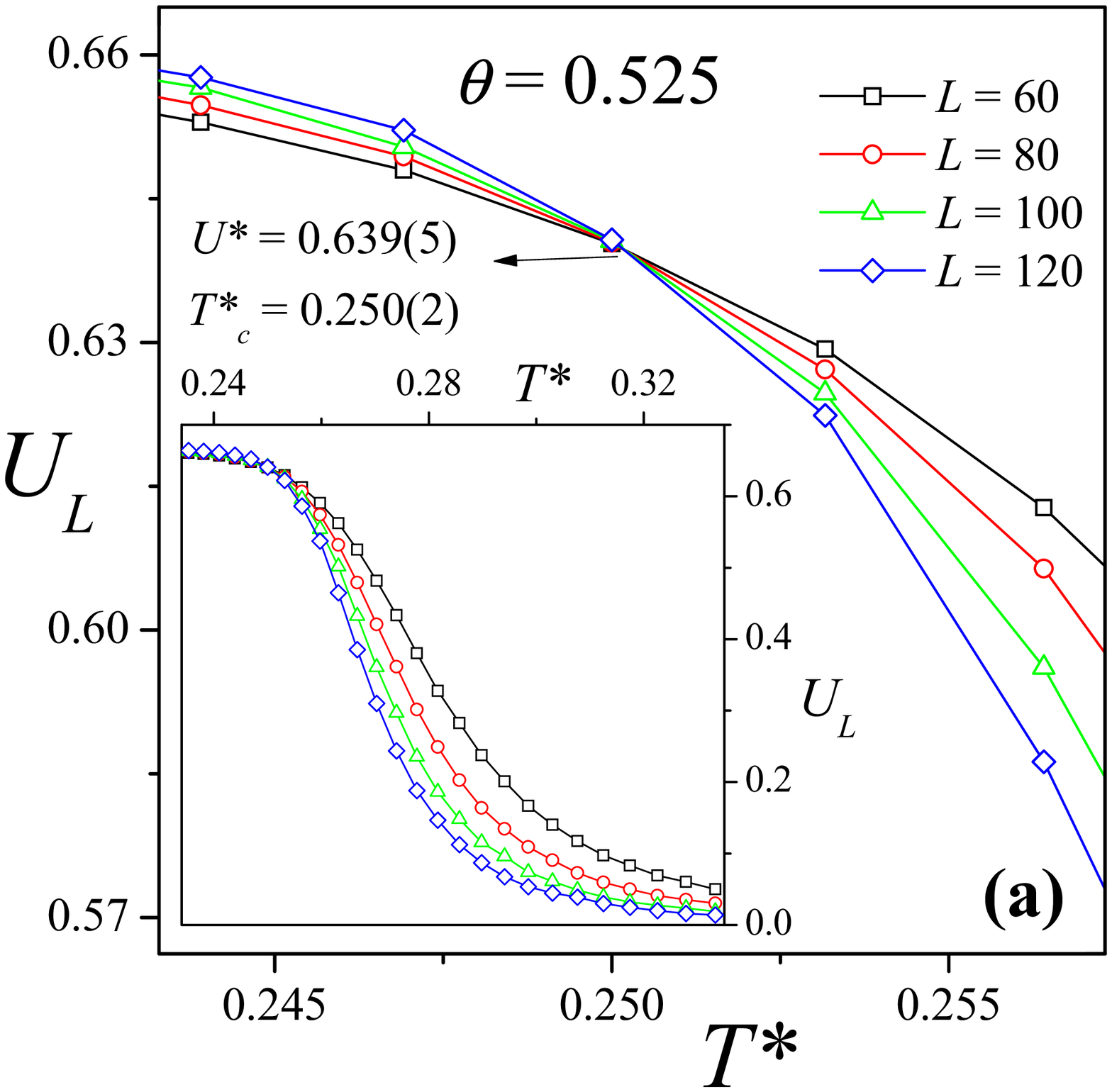}
\includegraphics[width=5cm,clip=true]{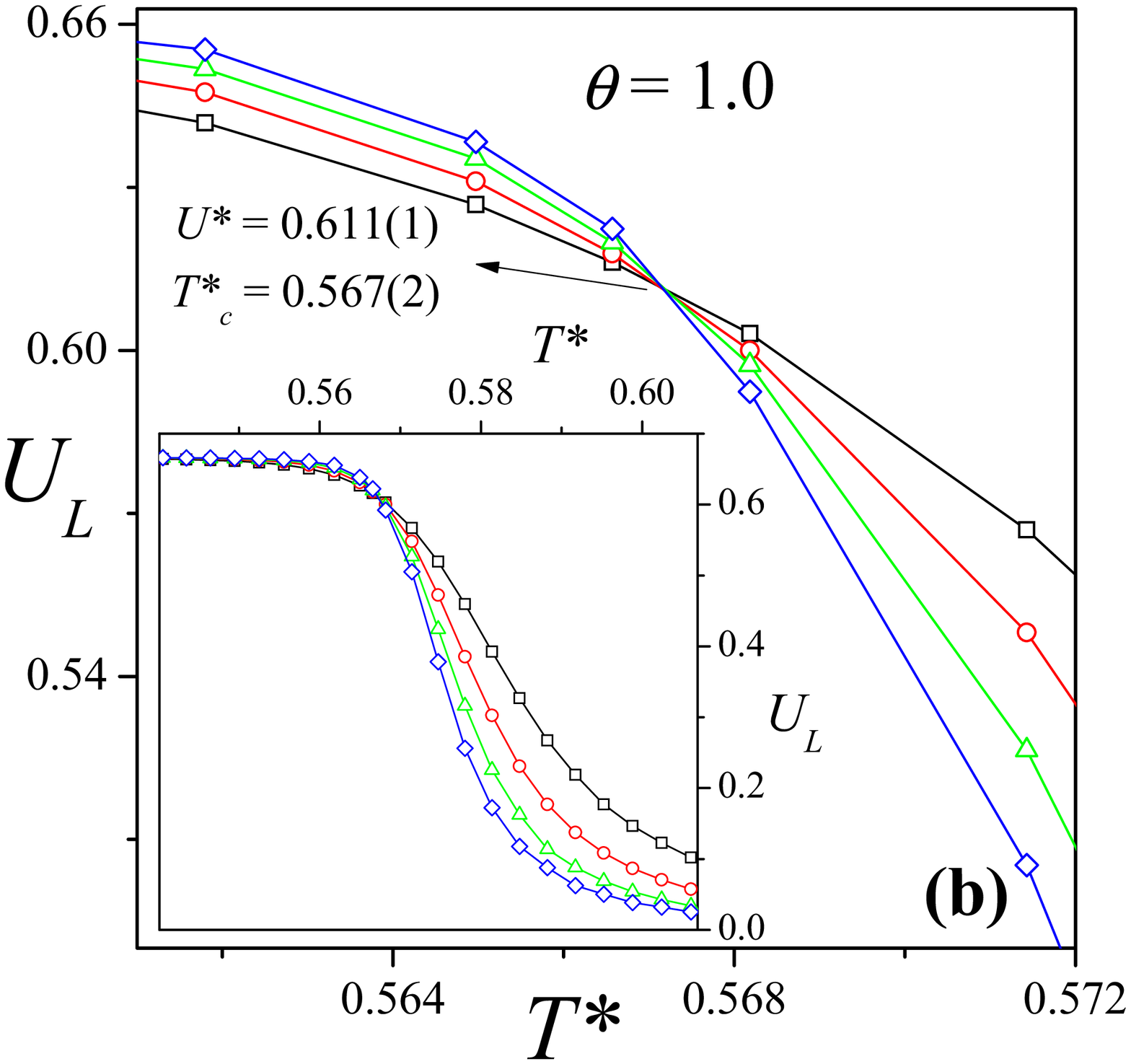}
\caption{Curves of $U_L$ vs $T^*$ for $\theta= 0.525$ (a) and
$\theta  = 1$ (b). From their intersections one obtained $T^*_c$.
In the insets, the data are plotted over a wider range of
temperatures. } \label{figure3}
\end{figure}

Hereafter we discuss the behavior of the critical temperature as a
function of coverage. The standard theory of finite-size scaling
allows for various efficient routes to estimate $T^*_c$ from MC
data \cite{BINDER,PRIVMAN}. One of these methods, which will be
used in this case, is from the temperature dependence of
$U_L(T^*)$, which is independent of the system size for
$T^*=T^*_c$. In other words, $T^*_c$ can be found from the
intersection of the curve $U_L(T^*)$ for different values of $L$,
since $U^* \equiv U_L(T^*_c)=$const. As an example, Fig. 3 shows
the reduced four-order cumulants $U_L$ plotted versus $T^*$ for
the cases studied in Fig. 2. The values obtained for the critical
temperature were $T^*_c = 0.250(2)$ (corresponding to
$\theta=0.525$) and $T^*_c = 0.567(2)$ (corresponding to
$\theta=1$). The procedure was repeated for $\theta$ ranging
between $0$ and $1$. The results, which are collected in Fig. 4
(a), represent the temperature-coverage phase diagram of the
system. The critical line (squares and line in the figure)
separates regions of isotropic and nematic stability. The
different phases are shown schematically in parts (b), (c) and (d)
of Fig. 4.

With respect to the numerical results obtained by Tavares et al.
at $\theta=0.2$ and $\theta=0.4$ [denoted with solid circles in
Fig. 4 (a)], the agreement with the present data is very good.

\begin{figure*}[t]
\includegraphics[width=6.5cm,clip=true]{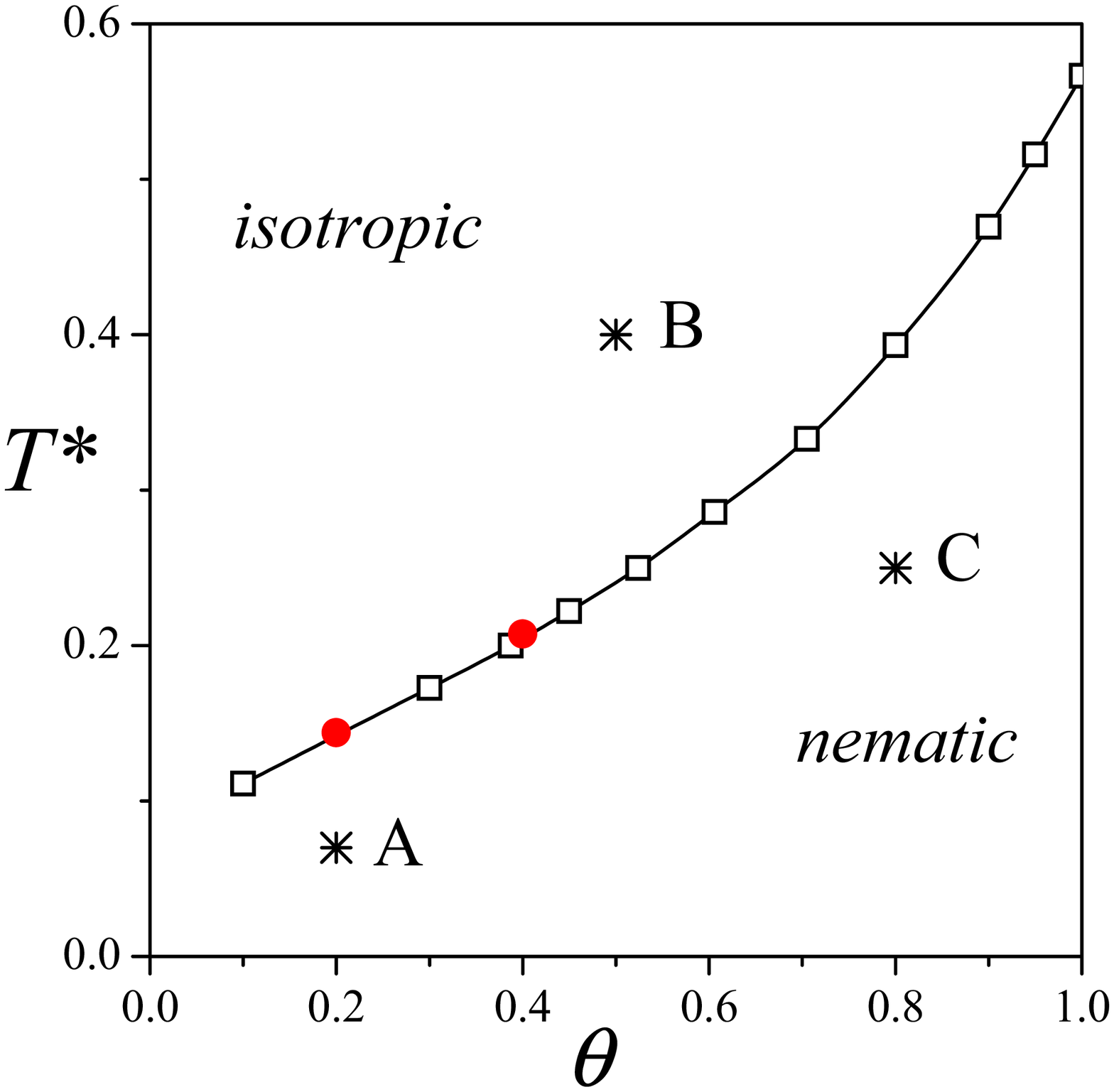}(a)
\includegraphics[width=6cm,clip=true]{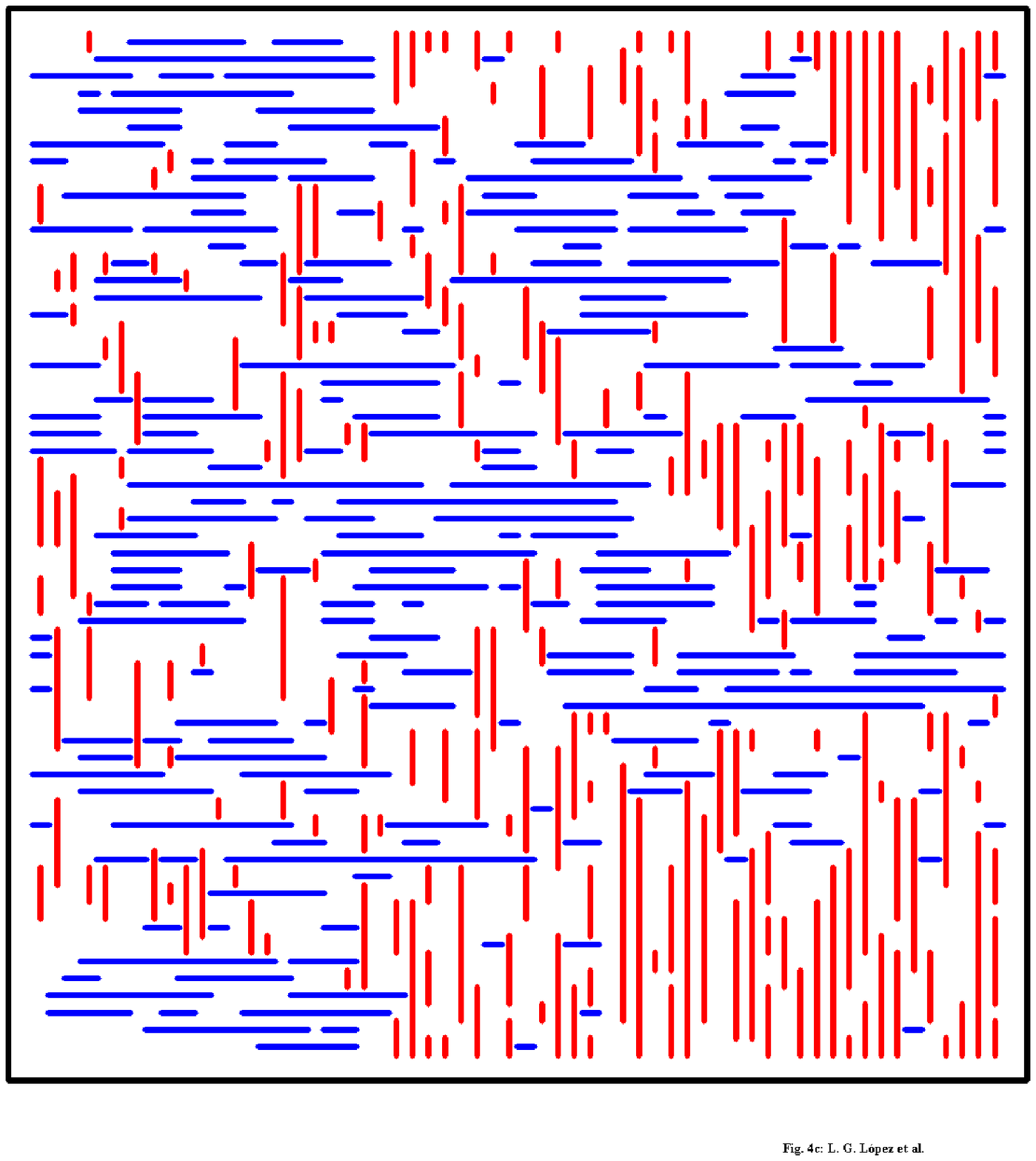}(c)
\includegraphics[width=6cm,clip=true]{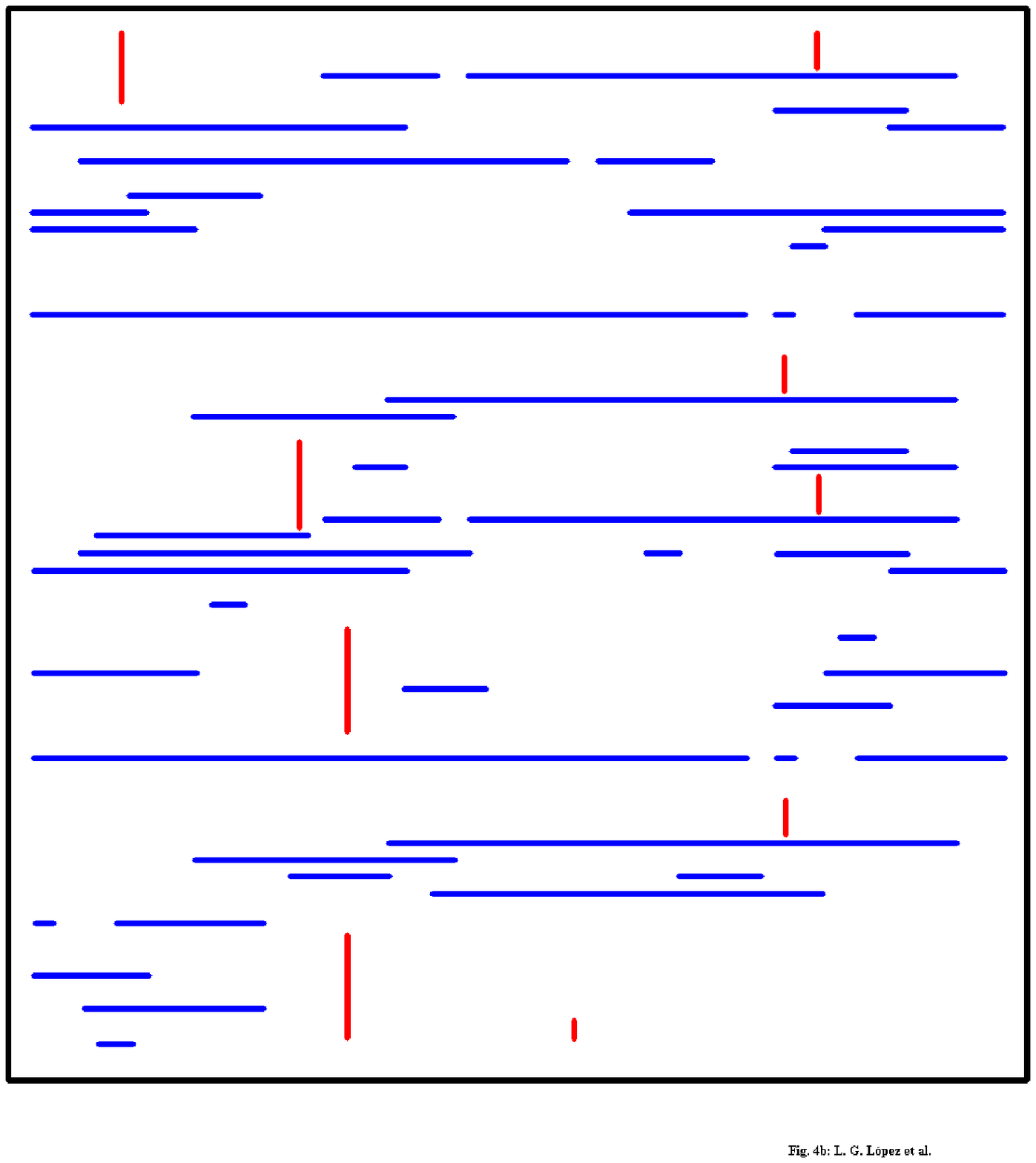}(b)
\includegraphics[width=6cm,clip=true]{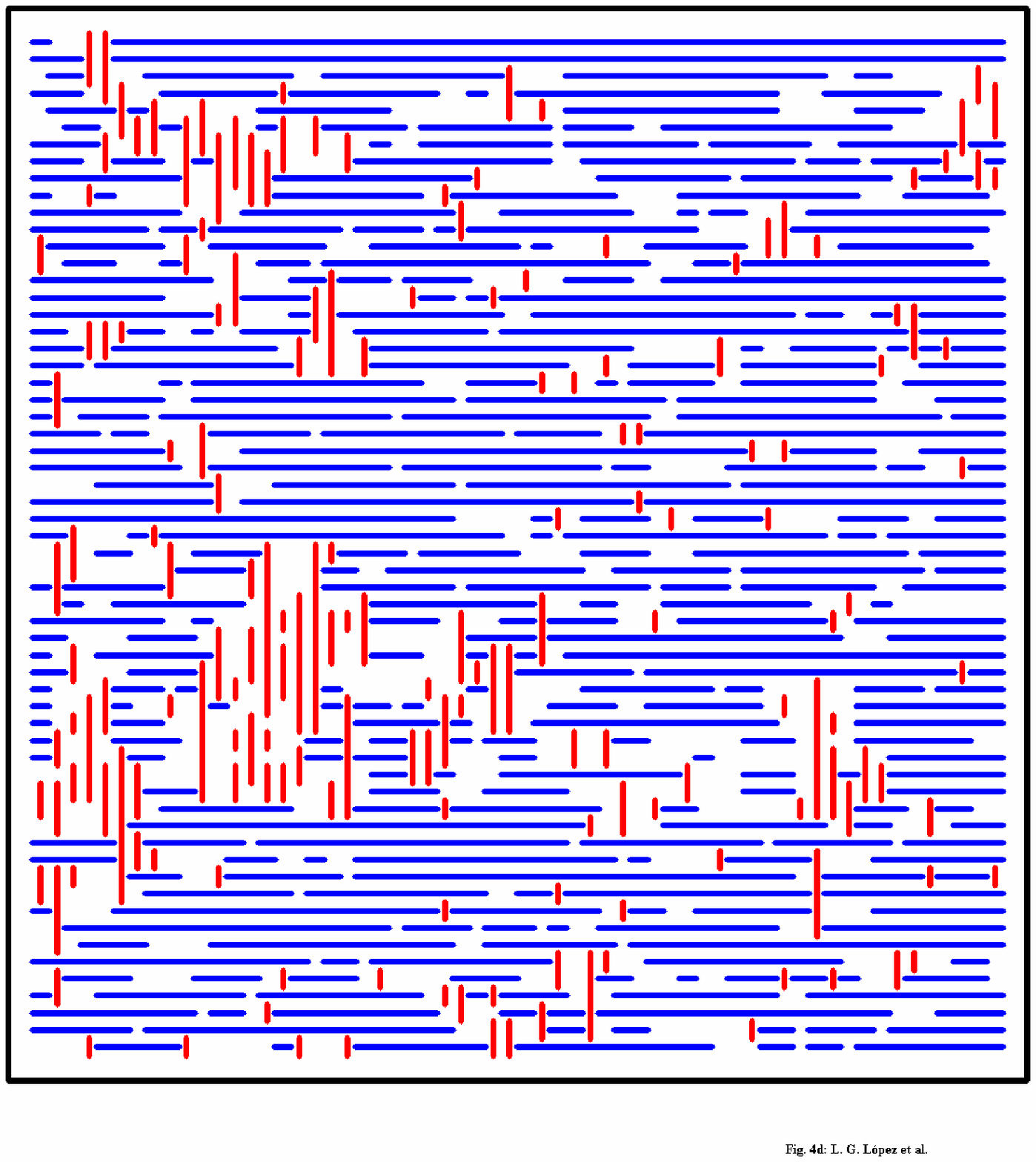}(d)
\caption{(a) Phase diagram of the model: our simulation data
(squares and line) and additional points (circles) obtained from
Monte Carlo simulation, carried out by Tavares et al.
\cite{Tavares}. (b) Schematic representation of the low-density
nematic phase (point A in the figure). (c) Same as (b) for the
intermediate-density disordered phase (point B in the figure). (d)
Same as (b) for the high-density nematic phase (point C in the
figure). } \label{figure4}
\end{figure*}

As it is well-known, the behavior of the reduced fourth-order
cumulant as a function of temperature not only provides an
accurate estimation of the critical temperature $T_c$ in the
infinite system, but also allows to make a preliminary
identification of the order and universality class of the phase
transition occurring in the system \cite{BINDER}. In the case of
Fig. 3, and as it is shown in the insets, the curves exhibit the
typical behavior of the cumulants in the presence of a continuous
phase transition. Namely, the order parameter cumulant shows a
smooth drop from $2/3$ to $0$, instead of a characteristic deep
(negative) minimum, as in a first-order phase transition
\cite{BINDER}.

With respect to the value of the intersection point $U^*$, two
different behaviors can be visualized from Fig. 3. On one hand, at
$\theta = 0.525$, the value obtained for $U^*$ ($U^* =0.639(5)$)
is consistent with the $q=1$ Potts universality class \cite{Wu}
observed in Ref. [\onlinecite{PRE4}], where the system was studied
at a fixed temperature ($T^* = 0.25$). On the other hand, and as
it is expected for $\theta=1$, the fixed value of the cumulants,
$U^* = 0.611(1)$, is consistent with the extremely precise
transfer matrix calculation of $U^* = 0.6106901(5)$ for the 2D
Ising model \cite{KAMIE}. Even though the value of $U^*$ may be
taken as a first indication of universality, a detailed
calculation of critical exponents is required for an accurate
determination of the universality class along the critical line in
Fig. 4 (a), and this will be subject of future research.

Finally, Fig. 5 shows the nematic order parameter $\delta$ as a
function of the coverage. The data correspond to $T^* = 0.25$ and
$L = 100$ \cite{foot3}. As the density is increased above a
critical value, the particles align along one direction and
$\delta$ increases continuously to one, remaining constant up to
full coverage. In other words, nematic order survives until
$\theta=1$. This finding $(1)$ allows us to discard the existence
of a reentrant nematic transition at high densities as speculated
in Ref. [\onlinecite{Tavares}] and $(2)$ indicates a substantial
difference between the present system and that of monodisperse
rigid rods without self-assembly, where a second nematic to
isotropic phase transition is observed at high densities
\cite{Ghosh,JSTAT1}.

\begin{figure}[t]
\includegraphics[width=6cm,clip=true]{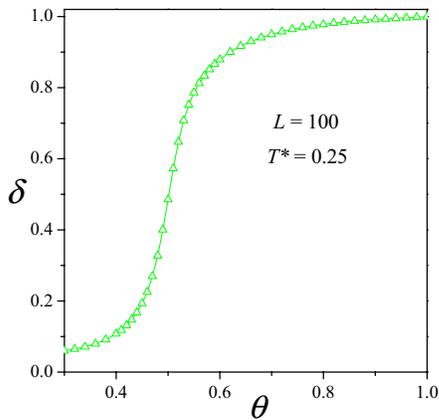}
\caption{Nematic order parameter $\delta$ as a function of the
coverage. The data correspond to $T^* = 0.25$ and $L = 100$. }
\label{figure5}
\end{figure}

\section{Analytical Approximations and Comparison Between
 Simulated and Theoretical Results}

In this section we calculate  the phase diagram  within mean field
and real space renormalization group approaches. Let

\begin{equation}\label{fgrand}
    f \equiv -\frac{1}{M\beta} \ln \left[{\rm Tr}\,  e^{-\beta H'} \right]
\end{equation}

\noindent the grand canonical free energy, where $H' =H-\mu N$,
$H$ and $N$ are given by Eqs. (\ref{H3}) and (\ref{Ns}), and $\mu$
is the chemical potential. The orientational order parameter and
coverage are then given by

\begin{equation}\label{orderp}
   \delta = \frac{1}{M} \sum_i \left< S_i \right> \, ,
\end{equation}

\noindent and

\begin{equation}\label{cov}
    \theta= \frac{1}{M} \sum_i \left< S_i^2 \right> \, ,
\end{equation}

\noindent respectively, where $\langle\cdots\rangle$ means here a grand canonical ensemble average.

\subsection{Mean-Field Approximation}

To obtain a mean field free energy $\Phi$ for this problem we use
the variational method \cite{ChLu2000}, based on Bogoliubov
inequality

\begin{equation}\label{Bogo}
    f \leq \Phi = f_0 + \frac{1}{M}\left<H'-H'_0 \right>_0,
\end{equation}

\noindent where $H'_0$ is a trial Hamiltonian containing variational parameters and
\[
f_0 =-\frac{1}{M\beta} \ln  \left[ {\rm Tr} \, e^{-\beta H'_0} \right]\,.
\]

\noindent We choose

\[
H'_0 = - \eta \sum_i S_i -\mu \sum_i  S_i^2,
\]

\noindent where $\eta$ is an effective field that breaks the orientational symmetry.  Then

\begin{equation}\label{Phi1}
    \Phi(\eta) = \eta\, \delta  - \frac{w}{2}\, (\delta^2+\theta^2) - \frac{1}{\beta}  \ln \left\{1 + 2\, e^{\beta\mu} \cosh (\beta\eta)
    \right\},
\end{equation}

\noindent where

\begin{equation}\label{m1}
     \delta = \left< S_i\right>_0=\frac{2\, e^{\beta\mu} \sinh (\beta\eta)}{1 + 2\, e^{\beta\mu}  \cosh
     (\beta\eta)},
\end{equation}
and
\begin{equation}\label{theta1}
    \theta= \left< S_i^2\right>_0= \frac{2\, e^{\beta\mu}\cosh (\beta\eta)}{1 + 2\, e^{\beta\mu} \ \cosh
    (\beta\eta)}.
\end{equation}

\noindent Minimizing Eq. (\ref{Phi1}) we obtain the self
consistent equation

\begin{equation}\label{CW}
     \eta= w\, \delta \left[1+\frac{\theta(1-\theta)}{\theta-\delta^2}
     \right].
\end{equation}

\noindent We see that  the isotropic state $\eta=0$ ($\delta=0$)
is a solution of Eq. (\ref{CW}). At low temperatures Eq.
(\ref{CW}) also presents  ordered (nematic) solutions $\eta\neq
0$. Making a Landau expansion of Eq. (\ref{CW})
 we obtain the following results:

\begin{itemize}
  \item There is a tricritical point at $ T^*_t = 3/4$ and $\mu_t = - \frac{3w}{4} \ln 2$, where $a_2=a_4=0$ and $a_6>0$.
  The coverage at this point is $\theta_t=1/2$.
  \item  When $\mu > \mu_t$ there is a second-order transition line  ($a_2=0$, $a_4>0)$ at

  \begin{equation}\label{criticalMF}
    \mu_c(T)= \frac{1}{\beta} \ln \left[\frac{1}{2}\left(\sqrt{\frac{\beta w}{\beta w -1}} -1
    \right)\right].
  \end{equation}

  \noindent Along the critical line we have $T^*_c= \theta(2-\theta)$.
  When $\mu \to \infty$ we have $\theta\to 1$ and $T^* \to 1$.
  Moreover, from Eqs. (\ref{m1}) and (\ref{CW}) we obtain in this limit $\delta = \tanh(\beta w \delta)$, i.e.,
  the mean field equation for an Ising model, as expected.

  \item When $\mu< -w$ only the isotropic solution remains. It is easy to see that there is a level crossing
  at $T^*=0$ between the empty state $\delta=\eta=0$ and the completely ordered one
  $\delta=\eta=1$.
  \item When $-w < \mu<\mu_t$ we have a first-order transition line ($a_4<0$ and $a_6>0$) which can be calculated numerically by a Maxwell construction.
\end{itemize}

In Fig. 6 we show the mean field phase diagram at the $(\mu,T^*)$
space, which is qualitatively similar to that of the isotropic
Blume-Emery-Griffiths (BEG) model \cite{BlEmGr1971}. The
corresponding phase diagram in $(\theta,T^*)$ space presents a
coexistence region between a low-coverage isotropic phase and a
high-coverage nematic one at low temperatures. The presence of
this coexistence region (first-order phase transition) is
completely at variance with the observed numerical simulation
results.

\begin{figure}[t]
\includegraphics[width=7cm,clip=true]{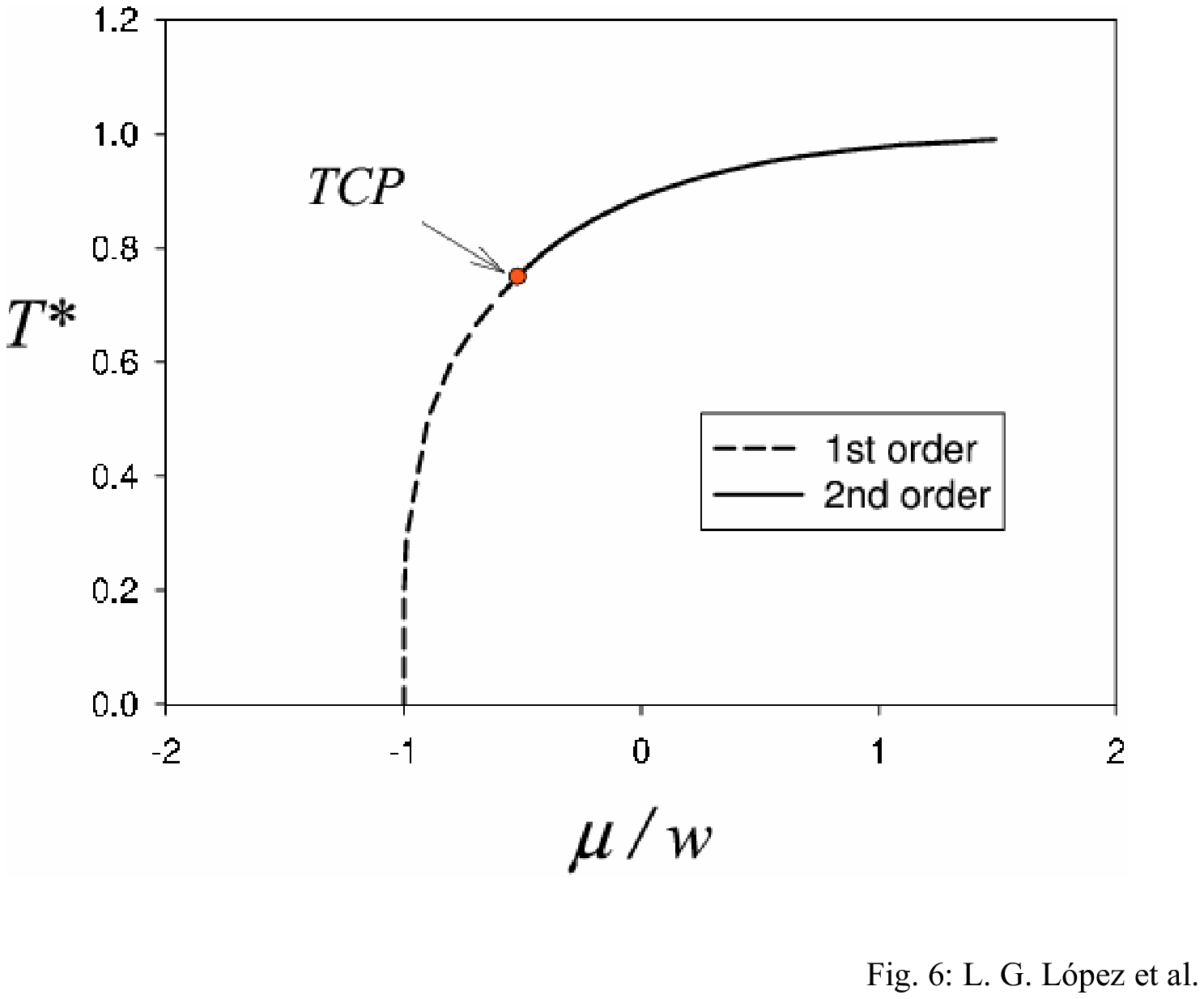}
\caption{$T^*$ vs. $\mu/w$ mean field  phase diagram. TCP is a
tricritical point. } \label{figure6}
\end{figure}

\subsection{Real Space Renormalization Group approach}

In order to obtain a more accurate analytical prediction for the
phase diagram we apply the Real Space Renormalization Group (RSRG)
scheme introduced by Niemeijer and van Leeuwen \cite{NiVa1982},
using  four spin Kadanoff blocks  and a double majority rule RG
projection matrix.  The details of the RSRG implementation are
given in the supplementary material \cite{SM}. The application of
a truncation scheme allowed us to restrict the proliferation of
interactions.  Under this framework, closed recursion RG relations
can be obtained for the the more general Hamiltonian compatible
with the basic symmetry of the system, namely, a 90 degrees
rotation of the lattice when $S_i \to -S_i$, that is
\begin{eqnarray}
    {\cal H}_{RG} & = &   h \sum_i S_i^2+ \sum_{<i,j>} \left[L\, S_i S_i + M\, S_i^2 S_j^2
    \right] \nonumber\\
    & & + \sum_{<i,j>} \left[ U\, (S_i^2 S_j + S_j^2 S_i)\left(\hat{y}.\vec{r}_{ij}-\hat{x}.\vec{r}_{ij}\right) \right] \label{HRG2}
\end{eqnarray}
\noindent where ${\cal H}\equiv -\beta H$ and $h \equiv \beta
\mu$. For $U=0$ the Hamiltonian (\ref{HRG2}) corresponds to the
BEG model \cite{BlEmGr1971}. For $L=M=U=\beta w/4$ we recover the
model (\ref{H3}).

The RG flow starting from the subspace
$(L,M,U,h)=(K/4,K/4,K/4,h)$, with $K\equiv \beta w$, is governed
by the following fixed points. {\it (i)} Two attractors at
$I_{\pm} = (0,0,0,\pm \infty)$. They represent the high
($\left<S_i^2 \right>\approx 1$) and the low ($\left<S_i^2
\right>\ll 1$) density isotropic phases respectively. {\it (ii)}
One semi-unstable fixed point $T_1 = (0,0,0,-\ln 2)$. It is the
locus of a surface in the $(L,M,U,h)$ space that corresponds to a
smooth continuation at high temperatures between both phases. {\it
(iii)} A line of attractive fixed points at
$(+\infty,0,0,+\infty)$. It is the locus of the  ferromagnetic
phase in the whole $(L,M,U,h)$ space and we  call it the $N$
attractor. {\it (iv)} One non-trivial fixed point
$C_1=(L_c,0,0,+\infty)$ with $L_c = \frac{1}{4}\, \ln \left[
1+2\sqrt{2}+ \sqrt{10+5\sqrt{2}}\, \right] \approx 0.518612$. It
is the locus of a critical surface and corresponds to the critical
point of the Ising model in the square lattice under the present
approximation. The associated critical exponent results
$\nu=1.0013...$, in excellent agreement with the exact result
$\nu=1$. The details of this analysis are given in the
supplementary material \cite{SM}.

\begin{figure}[t]
\includegraphics[scale=0.3,angle=-90]{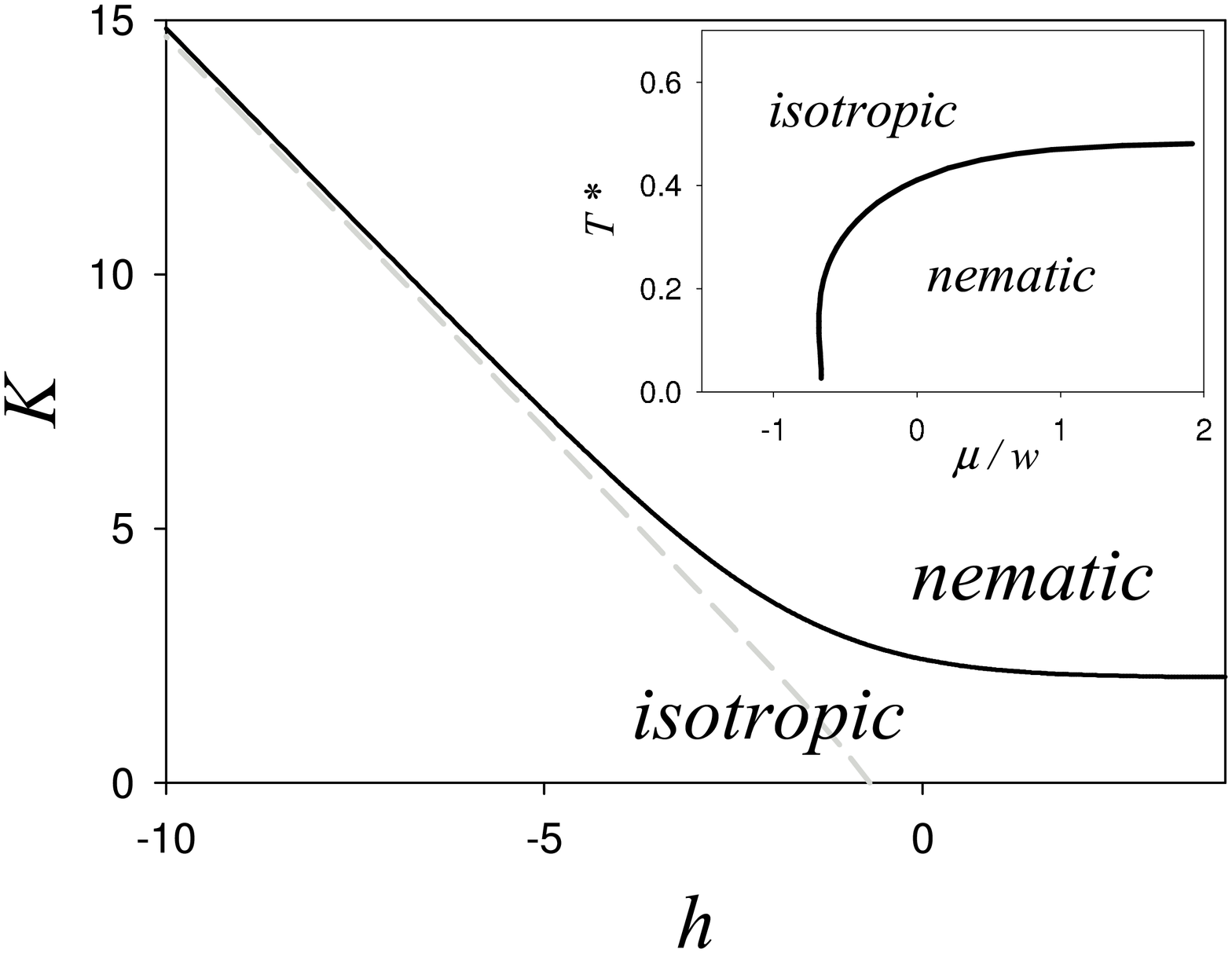}
\caption{RG phase diagram in the $(K,h)$ space. The black
continuous line is attracted by the fixed point $C_1$ and
therefore corresponds to a secondorder critical one. The grey
dashed line is attracted by the fixed point $T_1$ and corresponds
to a smooth continuation between the high and low density
isotropic phases. The inset shows the corresponding phase diagram
in the $(\mu/w, T^*)= (h/K,1/K)$ space. } \label{figure7}
\end{figure}

The phase diagram in the $(K,h)$ space,  obtained from the RG flow
starting with $(L,M,U,h)=(K/4,K/4,K/4,h)$, is shown in Fig. 7.  We
found a single critical line separating the nematic and isotropic
phases (black continuous line in Fig. 7), which is in the basin of
attraction of the fixed point $C_1$. The nematic phase is in the
basin of of attraction of $N$, while the isotropic phase is
attracted either by $I_+$ or by $I_-$. Points along the grey
dashed line in Fig. 7 are attracted by the trivial fixed point
$T_1$, thus corresponding to a smooth continuation from low to
high density isotropic phases, without phase transition. This line
converges asymptotically to the critical line when $h\to-\infty$.
Therefore, according to the present RG prediction the transition
is second order for any finite temperature and it is in the
universality class of the Ising model. The corresponding phase
diagram in the $(\mu/w, T^*)= (h/K,1/K)$ space is shown in the
inset of Fig. 7.

Finally, to calculate the phase diagram in the $(\theta,T^*)$
space we  computed numerically the coverage $\theta(h)$ along the
critical line $K_c=K_c(h)$ of Fig. 7. The results are presented in
Fig. 8 and the details of the calculation are given in the
supplementary material \cite{SM}.

\subsection{Comparison Between
 Theoretical and Simulated Results}

In Fig. 8 we compare the critical lines obtained by MC (open
squares joined by lines) and RSRG (solid circles), together with
the analytical  approximation developed by Tavares et al.
\cite{Tavares} (solid line). While qualitatively similar to the MC
result, we see that the present RSRG
 approximation systematically underestimates the critical temperature.
 Concerning the comparison with Tavares et al. results \cite{Tavares},
quantitative and qualitative differences have been found between
the analytical and the simulation data. In fact, the theory
overestimates the critical temperature in all range of coverage,
confirming the predictions in Ref. [\onlinecite{Tavares}]. For
small values of $\theta$, small differences appear between
simulation and theoretical results; however, the disagreement
turns out to be significantly large for larger $\theta$'s.

\begin{figure}[t]
\includegraphics[width=6cm,clip=true]{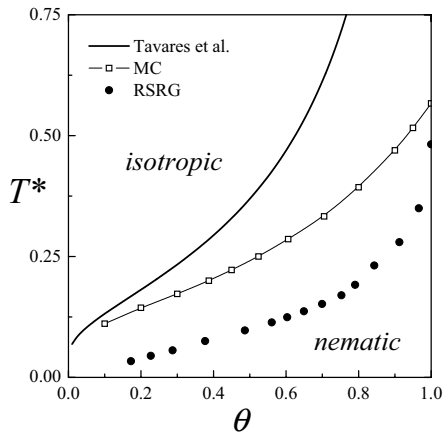}
\caption{Comparison between numerical and theoretical estimates of
the phase diagram in the $(\theta,T^*)$ phase diagram.}
\label{figure8}
\end{figure}

In the particular case of $\theta=1$, the Tavares et al. theory
predicts a critical temperature of $T^*_c=[\ln(3/2)]^{-1}
\approx 2.466$, whereas the value calculated by MC simulations is
$T^*_c = 0.567(2)$. These results can be compared with the
exact value of the critical temperature at full coverage $T^*_c
= -\left[2 \ln \left( \sqrt{2}-1 \right)\right]^{-1} \approx
0.567$, (see section \ref{Lattice-Gas Model}). This result is
consistent with that calculated by MC simulations, which
reinforces the robustness of the present computational scheme.

\section{Conclusions}

In summary, we have addressed the temperature-coverage phase
diagram of self-assembled rigid rods on square lattices. By using
Monte Carlo simulations, mean-field theory and a renormalization
group approach, we obtained and characterized the critical line
which separates regions of isotropic and nematic stability.
Several conclusions can be drawn from the present results.

First, a simulation test of the theory developed by Tavares et al.
\cite{Tavares} was carried out. The results showed that the theory
overestimates the critical temperature in all range of coverage,
confirming the predictions in Ref. [\onlinecite{Tavares}]. For
small values of $\theta$, small differences appear between
simulation and theoretical results; however, the disagreement
turns out to be significantly large for larger $\theta$'s. On the
other hand, the RSRG approach reproduces qualitatively the shape
of the critical line, but systematically underestimates the
critical temperature. Concerning this last calculation, the main
prediction is that the critical properties of the whole line are
associated to a unique second-order fixed point, confirming the
continuous nature of the transition. However, it must be pointed
out that it predicts that the whole line is in the universality
class of the $d=2$ ferromagnetic Ising model, at variance with
Monte Carlo numerical calculations  predicting that the transition
at $\theta \approx 1/2$ belongs to the $q=1$ Potts universality
class \cite{PRE4}. While the present RSRG results are not
conclusive, due to the approximate character of the approach, they
indicate that further research is required to clarify this point.

On the other hand, the behavior of the order parameter allowed to
discard the existence of a reentrant nematic transition at high
densities as speculated in Ref. [\onlinecite{Tavares}]. This
finding indicates a substantial difference between the present
system and that of monodisperse rigid rods without self-assembly,
where a second nematic to isotropic phase transition is observed
at high densities \cite{Ghosh,JSTAT1}

Concerning the MF results, the prediction of a first-order
transition line and a tricritical point is not surprising, due to
the close relationship between the present model and the BEG one,
as evidenced by the Eq. (\ref{H3}). Indeed, the generalized form
(\ref{HRG2}) contains both first-order and tricritical fixed
points, but the RSRG results show that in $d=2$ the anisotropic
character of the interactions drive the RG flow of the present
system outside their basins of attraction. However, in three
dimensional systems the IN transition is usually first-order
\cite{Onsager}. On the other hand, from the exact mapping into the
isotropic Ising model at full coverage one could expect a
second-order transition for high values of the coverage, even in
three dimensions. Hence, the MF prediction of a tricritical point
is probably correct for $d>2$.

\acknowledgments This work was supported in part by CONICET
(Argentina) under projects number PIP 112-200801-01332 and 112-200801-01576; Universidad
Nacional de San Luis (Argentina) under project 322000; Universidad Nacional C\'ordoba and the
National Agency of Scientific and Technological Promotion
(Argentina) under projects  PICT 2005 33328 and 33305 .



\section*{Supplementary information}

Here we provide the details of the RSRG calculations presented in
the manuscript.

\section*{RSRG Method}

Consider the Kadanoff blocks of size $N_b=b^2=4$ shown in Fig. 9.
Let's denote by $S_I'$ the block spin associated to the block $I$
and $s_I$ the set of lattice spins belonging to the block $I$:
$s_I \equiv \{ s_i\}\;\;$ with $\;\; i \in I$. Let's also denote
by $S'$ and $s$ the complete sets of block and lattice spins
respectively. We can express ${\cal H} = {\cal H}_0 + {\cal V}$,
where ${\cal H}_0 = \sum_I {\cal H}_I(s_I)$  contains all the
interactions between spins belonging to the block $I$ and $\cal V$
all the interblock interactions. Introducing an RG  projection
matrix $P(S',s)= \prod_I P_I(S_I',s_I)$, an average of an
arbitrary function $X(S',s)$ as

\begin{equation}
   \left<X \right>_0(S') \equiv  \frac{1}{Z_0}\;  \sum_{s} P(S',s) e^{{\cal H}_0(s)} X(S',s)
\end{equation}

\noindent where $Z_0 = \prod_I Z_0^I$ with

\[
 Z_0^I(S_I)= \sum_{s_I} P_I(S_I',s_I)  e^{{\cal H}_I(s_I)}
\]

\noindent The simplest RG approach within Niemeijer and van
Leeuwen \cite{NiVa1982} scheme consist into the identification

\begin{equation}\label{RG1}
    {\cal H}'(S') + {\cal C} = \ln Z_0+\left<V \right>_0,
\end{equation}

\noindent where ${\cal H}'(S')$ is the block Hamiltonian and
${\cal C}$ is a spin-independent constant. This uncontrolled
approximation results from the truncation to the first-order
cumulant\cite{NiVa1982} of $\left<\exp (V) \right>_0$. Using then
of the double majority rule RG projection matrix $P_I(S_I',s_I)$
introduced by Berker and Wortis for the pure isotropic
Blume-Emery-Griffiths (BEG) model \cite{BeWo1976}, it is easy to
see that $\left<S_{iI} \right>_0 = a_1\, S_I'$, $\left<S_{iI}^2
\right>_0 = a_2\, S_I'^2 + a_3$, and $\ln Z_0^I = a_4 S_I'^2 + a_5
$, where

\begin{eqnarray}
  a_1 &=& \left. \left<S_{iI} \right>_0\right|_{S_I'=1}\label{a1} \\
  a_2 &=& \left. \left<S_{iI}^2 \right>_0\right|_{S_I'=1}-\left.\left<S_{iI}^2 \right>_0\right|_{S_I'=0} \label{a2}\\
  a_3 &=& \left. \left<S_{iI}^2 \right>_0\right|_{S_I'=0}\label{a3}\\
  a_4 &=&\left. \ln Z_0^I\right|_{S_I'=1}-\left. \ln Z_0^I \right|_{S_I'=0}\label{a4} \\
  a_5 &=& \ln \left. Z_0^I \right|_{S_I'=0}.\label{a45}
\end{eqnarray}

\begin{figure}
\includegraphics[scale=0.3,angle=-90]{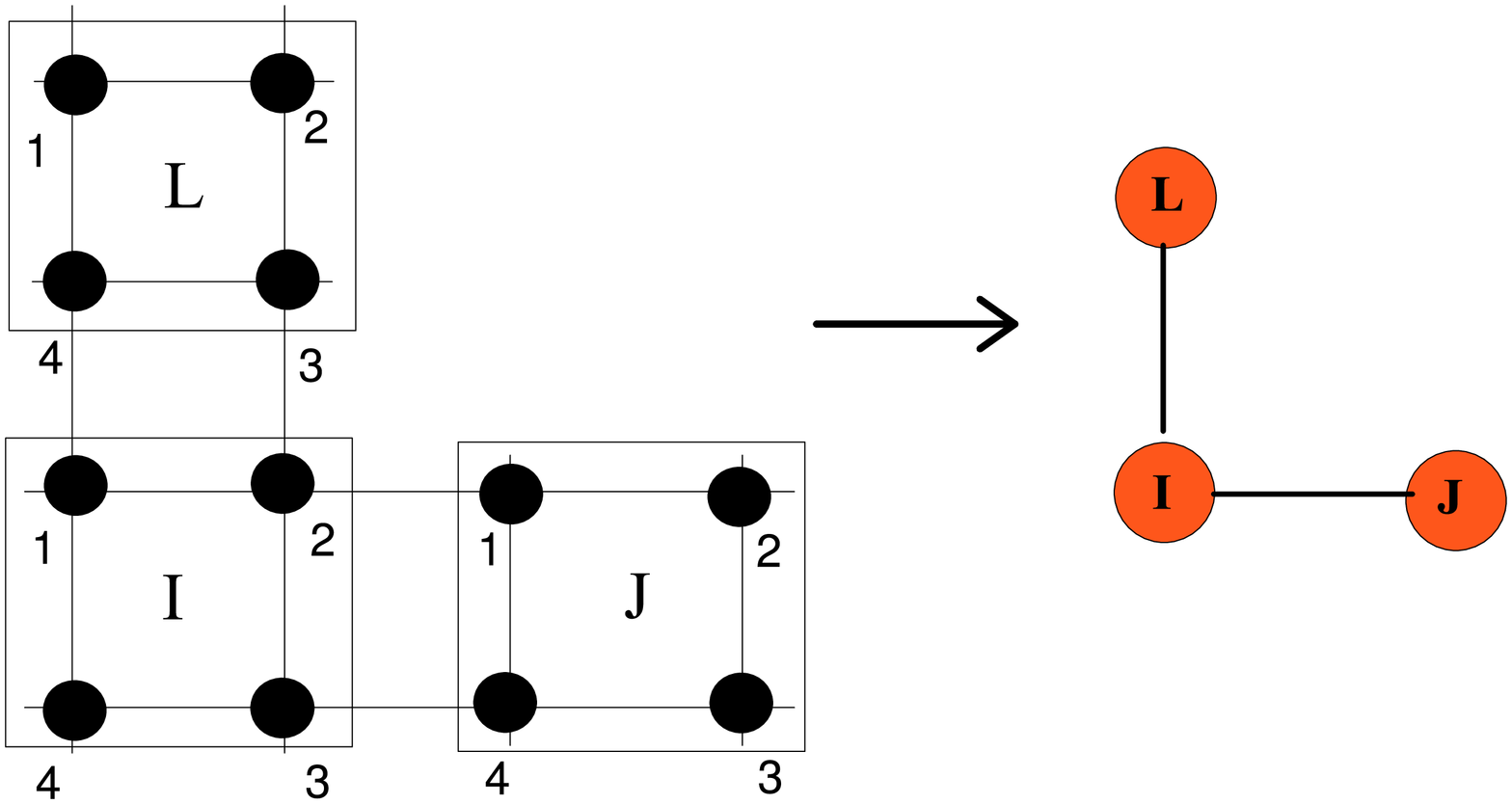}
\caption{ Kadanoff blocks of size $b=2$ for the square lattice. }
\label{figS1}
\end{figure}

Applying this scheme to the Hamiltonian

\begin{eqnarray}
    {\cal H}_{RG} &=&   h \sum_i S_i^2+ \sum_{<i,j>} \left[L\, S_i S_i + M\, S_i^2 S_j^2
    \right. \nonumber \\
    & & + \left. U\, (S_i^2 S_j + S_j^2 S_i)\left(\hat{y}.\vec{r}_{ij}-\hat{x}.\vec{r}_{ij}\right) \right], \label{HRG2}
\end{eqnarray}

\noindent we obtain the closed RG recursion relations

\begin{eqnarray}
  L' &=& 2L\, a_1^2 \label{RGEq1}\\
  M' &=& 2M\, a_2^2 \\
  U' &=& 2U\, a_1 a_2 \\
  h' &=& 8 M\, a_2\,a_3 + a_4, \label{RGEq4}
\end{eqnarray}

\noindent together with

\begin{equation}\label{CC}
    g= {\cal C}/N= ( M\, a_3^2  + a_5/4).
\end{equation}

\noindent Defining

\begin{eqnarray}
  B_1(L,M,U,h) &=& \left. Z_0^I \right|_{S_I'=0}\nonumber \\
  B_2(L,M,U,h) &=& \left. Z_0^I \right|_{S_I'=1}\nonumber
\end{eqnarray}

\noindent we obtain

\begin{eqnarray}
  a_3 &=& \frac{2\, e^{h}+ 2\, e^{2h} + 2\, e^{2h+M-L} + 2\, e^{2h+M+L} \cosh(2U) }{B_1(L,M,U,h)}\nonumber\\
  a_4 &=& \ln \frac{B_2(L,M,U,h)}{B_1(L,M,U,h)}\nonumber \\
  a_5 &=& \ln B_1(L,M,U,h)\nonumber
\end{eqnarray}

\noindent and

\begin{widetext}
\begin{eqnarray}
  a_1 &=& \frac{1}{B_2(L,M,U,h)} \left[ \frac{1}{2}\, e^{2h} + 2\,e^{4(h+M)}+ e^{3h+2(M-L)} + 3\,  e^{3h+ 2(M+L)}+ e^{4(h+ M+L)} + \right. \nonumber\\
  &+& \left. e^{2h+M+L}\, \cosh (2U) + 2\, e^{3h+2M}\, \cosh (2U)\right] \nonumber\\
  a_2 &=& \frac{1}{B_2(L,M,U,h)} \left[ e^{2h} + 6\, e^{4(M+h)}+  e^{2h +M-L} + 6\,  e^{3h+2M}\,\cosh(2L) + 2\, e^{4(M+h)}\,\cosh(4L) + \right. \nonumber \\
    &+&  \left. e^{2h+M+L}\,\cosh(2U) +  6\, e^{2M+3h}\,\cosh (2U)\right] - a_3(L,M,U,h) \nonumber\\
  a_3 &=& \frac{2\, e^{h}+ 2\, e^{2h} + 2\, e^{2h+M-L} + 2\, e^{2h+M+L} \cosh(2U) }{B_1(L,M,U,h)} \nonumber
\end{eqnarray}

\begin{eqnarray}
  B_1 &=& 1+ 8\, e^{h}+ 4\, e^{2h} +4\, e^{2h+M-L} + 4\, e^{2h+M+L}\, \cosh (2U)\nonumber \\
  B_2 &=& 2\, e^{2h}  + 6\, e^{4(M+h)}+ 2\, e^{2h+M-L} + 8\, e^{3h+2M}\, \cosh(2L) + 2\,  e^{4(M+h)}\,\cosh(4L) + \nonumber \\
     &+&  2\,  e^{2h+M+L} \,\cosh(2U) +8\,  e^{3h+2M} \,\cosh(2U). \nonumber
\end{eqnarray}
\end{widetext}

\section*{RSRG flow and fixed points structure}

We found that all the relevant fixed points of the recursion
equations lie in the BEG subspace $U=0$. The RG flow and the fixed
points structure in the $U=0$ subspace is qualitatively similar to
that obtained in Ref. [\onlinecite{BeWo1976}], including first and
second-order surfaces, as well as  tricritical and critical
endpoint lines \cite{BeWo1976}. We  focused only on those fixed
points relevant to the present problem namely,  those which govern
the RG flow starting from the subspace
$(L,M,U,h)=(K/4,K/4,K/4,h)$, with $K\equiv \beta w$.
 The whole
flow starting from that subspace is attracted by two  subspaces
invariant under RG: $L=M=U=0$ and $(M,h)=(0,+\infty)$.

\subsection*{Flow in the $L=M=U=0$ subspace}

 The recursion relations in this case reduce  to $h'= a_4(0,0,0,h)$.
  This RG equation  presents three fixed points: two attractors at $h=\pm \infty$,
  which are the loci of the high ($h=\infty$) and low ($h=-\infty$) density isotropic phases respectively and one unstable high temperature fixed point at $h= -\ln 2$. The first two fixed points are attractors in the complete $(L,M,U,h)$ space and we will call them $I_+$ and $I_-$. They represent the high ($\left<S_i^2 \right>\approx 1$) and the low ($\left<S_i^2 \right>\ll 1$) density isotropic phases respectively. The  fixed point $T_1 \equiv (0,0,0,-\ln 2)$ is the locus of a surface in the $(L,M,U,h)$ space that corresponds to a smooth continuation at high temperatures between both phases.

\subsection*{Flow in the $(M,h)=(0,+\infty)$ subspace}

 This subspace corresponds to an anisotropic Ising model, since in this limit the $S_i=0$ state has zero probability. The recursion relations reduce in this case to

\begin{eqnarray}
  L' &=& 2L\, d(L)^2 \\
  U' &=& U d(L)
\end{eqnarray}

\noindent with

\begin{equation}\label{dl}
    d(L) = \lim_{h\to\infty} a_1(L,0,U,h)= \frac{2+e^{4L}}{6+2\,
    \cosh(4L)}.
\end{equation}

\noindent Since $L'=L'(L)$, independently of the parameter $U$,
the whole flow is governed by the RG equation corresponding to the
isotropic Ising model. This equation has a non trivial fixed point
at $d(L) =1/\sqrt{2}$, whose solution is $L_c = \frac{1}{4}\, \ln
\left[ 1+2\sqrt{2}+ \sqrt{10+5\sqrt{2}}\, \right] \approx
0.518612$, corresponding to the critical point of the Ising model
in the square lattice under the present approximation (compare
with exact Onsager result $L_c=\frac{1}{2} \ln(1+\sqrt{2})\approx
0.44069$). We will call this fixed point $C_1$. The critical
exponent $\nu$ is given by

\[
\nu =\frac{\ln b}{\ln \lambda} \;\;\;\; \lambda=\left.
\frac{\partial L'}{\partial L}\right|_{L_c}.
\]

\noindent We obtain $\nu=1.0013...$, in excellent agreement with
the exact result $\nu=1$. The RG recursion equation also has two
attractors: $I_+$ ($L=0$) and the isotropic ferromagnetic fixed
point $L=\infty$ ($T=0$). At $L=L_c$ we have another invariant
line at the $(L,U)$ space, whose RG equation is $U'= U'/
\sqrt{2}$.  This recursion relation has only trivial fixed points:
one attractor at $U=0$ and one unstable at $U=+\infty$. The line
$L=0$ is also invariant and have the same fixed points. Finally,
we have that $\lim_{L\to\infty} d(L)=1$
 Hence,   $U'=U$ and the whole line $L=+\infty$ is a line of fixed points.
This is the locus of the  ferromagnetic phase in the whole
$(L,M,U,h)$ space and we will call it the $N$ attractor. In Fig.
10 we show the flow diagram in the complete $(U,L)$ space.

\begin{figure}
\includegraphics[scale=0.3,angle=-90]{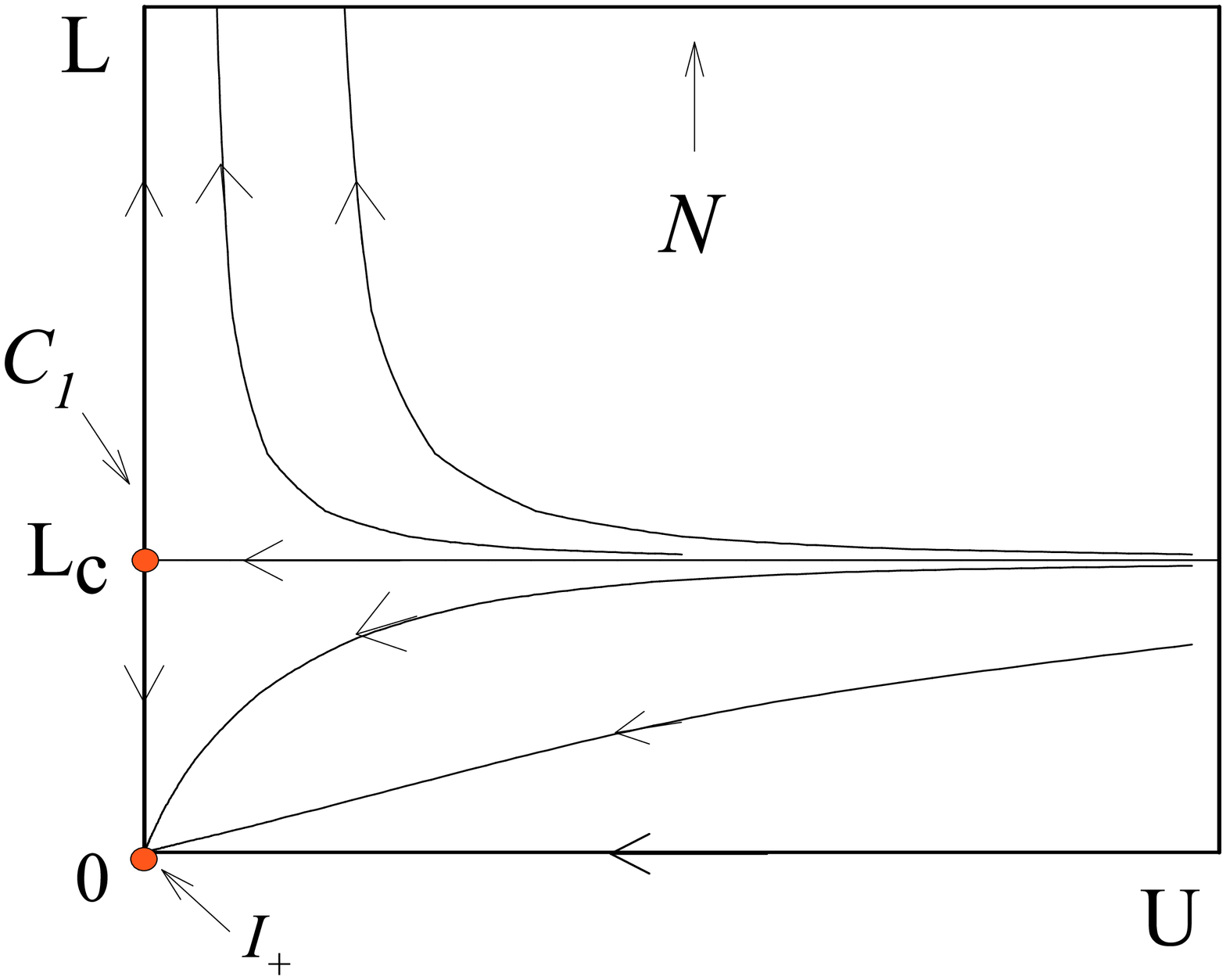}
\caption{ RG flow in the $(L,0,U,+\infty)$ invariant subspace. }
\label{figS2}
\end{figure}

\section*{RSRG Coverage calculation}

The coverage can be expressed as

\begin{equation}\label{thetarg}
   \theta(K,h) = -\beta \frac{\partial f(K,h)}{\partial h}.
\end{equation}

\noindent  Let $\vec{K}\equiv(L,M,U,h)$ be the parameters vector
of Hamiltonian (\ref{HRG2}). From the renormalization group
transformation we have the following relation after $n$
applications of the RG transformation \cite{NiVa1982}

\begin{equation}\label{frg}
f(\vec{K}_0) = -\frac{1}{\beta}\sum_{m=0}^n b^{-md} g(\vec{K}_m) +
b^{-nd} f(\vec{K}_n),
\end{equation}

\noindent where $\vec{K}_m$ is the parameters vector after $m$
applications of the RG transformation, $\vec{K}_0$ is the initial
value and $g(\vec{K})={\cal C}/N$ is given by Eq. (\ref{CC}).
Since $\theta$ is not singular at the critical line, we can assume
that the singular part of the free energy will make no
contribution to Eq. (\ref{frg}) and therefore  the derivative of
the second term in the right hand of the previous expression
vanishes when $n\to\infty$. Therefore, we can express

\begin{equation}\label{thetarg2}
   \theta(K,h) = \frac{\partial}{\partial h}\left[\sum_{m=0}^\infty b^{-md}
   g(\vec{K}_m)\right]_{\vec{K}_0=(K/4,K/4,K,4,h)}.
\end{equation}

\noindent Computing numerically the above sum and taking the
numerical derivative we obtain the critical line $T^*$ vs.
$\theta$ shown in Fig. 8 of the manuscript.


\begin{thebibliography}{99}
\bibitem{foot1} In Ref. \cite{PRE4}, we study the critical
behavior of the present model at $T^* = 0.25$ and $\theta \approx
0.525$. There, we set the reduced temperature to $T^* = 0.25$ and
performed a finite-size scaling analysis in terms of density,
obtaining $\theta_c \approx 0.525$. In the present paper, and in
order to corroborate the previous results, we repeat the scaling
treatment, this time maintaining constant the surface coverage (at
$\theta = 0.525$) and varying the temperature of the system.
\bibitem{foot3} In this case, we set the temperature $T$, varied the
density $\theta =N/M$ and monitored the order parameter $\delta$.
\end{thebibliography}
\end{document}